\documentclass[aps,prl,twocolumn,superscriptaddress,showpacs,showkeys,
notitlepage]{revtex4-1}

\usepackage{times,amsmath,amsfonts,amssymb,mathrsfs,graphics,graphicx,color,
comment,bm}
\usepackage[next]{inputenc}
\usepackage[dvips]{epsfig}
\usepackage[pdfstartview=FitH]{hyperref}
\usepackage{natbib}
\usepackage{pdfsync}
\usepackage{appendix}

\begin{document}

\title{Antiferromagnetism and Hidden Order in Isoelectronic Doping of 
URu$_2$Si$_2$}

\author{M.N. Wilson}
\address{Department of Physics and Astronomy, McMaster University, Hamilton,
 Ontario L8S 4M1, Canada}
\author{T.J. Williams}
\address{Quantum Condensed Matter Division, Neutron Sciences Directorate, Oak 
Ridge National Laboratory, Oak Ridge, TN 37831, USA}
\author{Y.-P. Cai}
\address{Department of Physics and Astronomy, McMaster University, Hamilton,
 Ontario L8S 4M1, Canada}
\author{A.M. Hallas}
\address{Department of Physics and Astronomy, McMaster University, Hamilton,
 Ontario L8S 4M1, Canada}
\author{T. Medina}
\address{Department of Physics and Astronomy, McMaster University, Hamilton,
 Ontario L8S 4M1, Canada}
\author{T.J. Munsie}
\address{Department of Physics and Astronomy, McMaster University, Hamilton,
 Ontario L8S 4M1, Canada}
\author{S.C. Cheung}
\address{Department of Physics, Columbia University, New York, New York 10027,
 USA}
\author{B.A. Frandsen}
\address{Department of Physics, Columbia University, New York, New York 10027,
 USA}
\author{L. Liu}
\address{Department of Physics, Columbia University, New York, New York 10027,
 USA}
\author{Y.J. Uemura}
\address{Department of Physics, Columbia University, New York, New York 10027,
 USA}
\author{G.M. Luke}
\address{Department of Physics and Astronomy, McMaster University, Hamilton,
 Ontario L8S 4M1, Canada}
\address{Canadian Institute for Advanced Research, Toronto, Ontario M5G 1Z7,
 Canada}

\begin{abstract}
We present muon spin rotation ($\mu$SR) and susceptibility measurements on 
single crystals of isoelectronically doped URu$_{2-x}$T$_x$Si$_2$ (T = Fe, Os)
 for doping levels up to 50\%. Zero Field (ZF) $\mu$SR measurements show
 long-lived oscillations demonstrating that an antiferromagnetic state exists
 down to low doping levels for both Os and Fe dopants. The measurements 
further show an increase in the internal field with doping for both Fe and Os. 
Comparison of the local moment - hybridization crossover temperature from
 susceptibility measurements and our magnetic transition temperature shows 
that changes in hybridization, rather than solely chemical pressure, are
 important in driving the evolution of magnetic order with doping.
\end{abstract}

\maketitle
\section{Introduction}

Heavy fermion systems frequently exhibit interesting electronic ground states 
arising from complex hybridization between conduction electrons and localized
 $f$-electrons\cite{Stewart1984}. Compounds containing uranium are particularly
 interesting as the Coulomb interaction, spin-orbit coupling, and 5f electron
 bandwidth are all of comparable energies, making exotic ground states possible
 \cite{Mydosh2011}. A notable example of such a ground state is the 
`Hidden Order' (HO) arising in URu$_2$Si$_2$ below $T_0 = 17.5$~K that
 was first studied in 1985 \cite{Palstra1985, Schlabitz1986}. The order in this
 state is termed `Hidden' as, despite more than two decades of study, the order
 parameter for the 17.5~K transition has not yet been conclusively identified
 \cite{Mydosh2011}.

Early neutron scattering studies indicated that this state was antiferromagnetic
 with a moment of 0.02-0.04~$\mu_B$ per uranium\cite{Broholm1987, Isaacs1990}. 
However, other studies found unusual properties that could not be explained by
 simple antiferromagnetism, such as a gap opening up over a large portion of
 the Fermi surface indicated by specific heat \cite{Maple1986} and infrared
 spectroscopy \cite{Bonn1988} measurements. Furthermore, the measured
 antiferromagnetic moment  is too small to explain the 0.2$R$ln2 per f.u.
entropy change across the transition determined from specific heat measurements
 \cite{Maple1986}.

Subsequent neutron scattering measurements conducted under applied hydrostatic
 pressure demonstrated a first-order transition into a large moment 
antiferromagnetic state (LMAF) with a moment of 0.4 $\mu_B$\cite{Amitsuka1999}
 that occurs at a critical pressure of 0.5-0.8~$GPa$ \cite{Butch2010PRB}.
 $\mu$SR measurements under applied pressure have also confirmed this
 first-order transition to the LMAF state, and demonstrate no
pressure-dependence of the internal fields from 0.5-1.5~$GPa$ \cite{Amato2004}.
  In addition, $\mu$SR \cite{MacLaughlin1988, Luke1994} and NMR measurements
 \cite{Matsuda2001} show that the weak antiferromagnetic moment seen at ambient
 pressure can be explained by a small phase separated volume fraction of the 
pressure-induced antiferromagnetic state coexisting with the hidden order state.
 It is now widely accepted that this low moment antiferromagnetism is extrinsic
 to the hidden order state and is caused by inhomogeneous strain in measured
 crystals \cite{Amitsuka2007}. 

The origin of the entropy change in URu$_2$Si$_2$ seen from heat capacity
 measurements has recently been explained by a gap opening in the spin
 excitation spectrum at the transition, and does not require the presence of
 weak antiferromagnetism \cite{Wiebe2007}. This gap is equivalent to the Fermi
surface becoming gapped, and angle-resolved photo-emission spectroscopy
 (ARPES) measurements \cite{Syro2009, Chatterjee2013} indicate that this gap
 arises from hybridization of the conduction band with the uranium
 5$f$-electrons. Scanning
 tunneling microscopy (STM) measurements \cite{Schmidt2010} have lent support
 to this idea by observing a band splitting below the hidden order transition.
 However, these results have been disputed, with other STM researchers
 \cite{Syro2009} claiming that the hybridization gap opens well above $T_{HO}$
 and hence cannot explain the hidden order state. This leaves the importance of
the hybridization gap as one of the many unanswered questions of URu$_2$Si$_2$.

Despite these significant advances in the understanding of HO a viable theory
 has not yet been accepted to explain this state, although numerous theories
 have been advanced over the years (See Ref. \cite{Mydosh2014} for a recent
 overview). In order to constrain such theories it is advantageous to further
 study the hidden order state through various experiment perturbations. One
 such perturbation that has been extensively applied to URu$_2$Si$_2$ is
 chemical doping. Previous studies have found that doping of the silicon site
 has only a weak effect on the electronic state which may be explained by a
 chemical pressure effect\cite{Dhar1992, Park1994-1}, while doping of the
 uranium \cite{Ocko1997, Amitsuka2000} and ruthenium
 \cite{Amitsuka1988, Dalichaouch1990PB, Dalichaouch1990PRB, Park1994-2}
 sites cause much more dramatic changes in the behavior. This indicates that
 the electronic ground state depends much more strongly on $d$-$f$ electron
 hybridization than it does on $sp$-$f$ hybridization \cite{Park1994-1}.
 However, U-site doping is complicated as there is competition between dilution
 of the magnetic U atom, changes in lattice parameters, and hybridization all
 occurring with doping. This makes Ru-site doping interesting to study as it is
a potentially simpler avenue to explore the effect of changing hybridization on
 the magnetic states. 

Rhodium and rhenium doping are two cases that have been well studied, both
 of which suppress the HO state before 5\% doping. However, the ground states
 that emerge after the suppression are distinctly different. For Re doping the
 HO transition is suppressed by a 5\% doping level, and above 7.5\% doping a
 non-Fermi liquid ferromagnetic state emerges that persists up to high doping
 levels \cite{Butch2010}. By contrast, Rh suppresses HO by 2\% doping at which
 point a LMAF state emerges, which is in turn suppressed by 4\% doping 
\cite{Yokoyama2004}. Above this doping level no magnetically ordered state is
observed \cite{Yokoyama2004}.

The Rh doped system has been a particularly valuable avenue to study the 
competition between the LMAF and HO states in the URu$_2$Si$_2$ system,
 as the doping allows the transition to be studied without the experimentally 
challenging aspects of applied external pressure. This has allowed productive
 studies of the high field behavior of the HO state (See Ref. \cite{Mydosh2011}
 and references therein), as well as proposed identification of universal
parameters that cause the transition between the HO and LMAF states
\cite{Yokoyama2004}.

Despite the potential insights gained by studying Re and Rh doping, the
 interpretation of results from both of these systems is made more difficult
 because these dopings change multiple potentially important parameters 
simultaneously. In particular, doping of Re or Rh will change the number of 
electrons in the system, the $d$-$f$ hybridization, and the lattice constants 
of the system. In order to more easily understand the mechanisms behind the
 transitions between HO and other phases it is beneficial to have systems that
 change as few parameters as possible in order to isolate their effects. This
 makes the isoelectronic dopings, osmium and iron, interesting to study as
 one does not have to consider the effect of changing electron numbers in 
this system.

Fe doping of URu$_2$Si$_2$ has been studied for polycrystalline samples by
 Kanchanavatee \textit{et al.} \cite{Kanchanavatee2011}. This work demonstrated
that the full range of compositions URu$_{2-x}$Fe$_x$Si$_2$ from x = 0 to 2 can
 be produced, and that doping results in a monotonic decrease of the lattice
 parameters with no evidence for a change of structure. Furthermore, the
 temperature-doping phase diagram measured by bulk probes (specific heat,
 magnetization, and resistivity) shows an increase in transition temperature as
 a function of doping up to a maximum of 40~K. This increase parallels that of
 the pure compound under pressure, which led the authors to hypothesize a
 transition from HO to LMAF at a doping level of $x = 0.1$ and conclude that
the effect of Fe doping on the system is fully explained by a chemical pressure
 effect \cite{Kanchanavatee2011}. However, the LMAF and HO states are largely
indistinguishable to the bulk probes used in this study and the authors did not 
perform measurements with any microscopic probes that would allow the
 magnetic state to be identified, hence no firm conclusions could be drawn.

Very recently, a second study has been published on Fe doping using neutron
 diffraction on single crystals \cite{Das2015}. In this work, elastic neutron
 scattering allowed the authors to identify a crossover from HO to AF at a
 doping level of $x = 0.1$ as would be expected from a chemical pressure
argument. However, the moment of 0.8~$\mu_B$ per U that they observe is twice
that seen in the pure material under pressure which indicates that chemical
pressure is not the only factor governing the evolution of magnetism in this
material. This discrepancy makes further study of Fe doping valuable to properly
understand the HO to LMAF transition if it is to be used as an analogue of the
pressure induced transition.

A cursory study of polycrystalline URu$_{2-x}$Os$_x$Si$_2$ was first performed
by Dalichaouch \textit{et al.} in 1990 \cite{Dalichaouch1990PRB}, and has been
recently followed by a more detailed examination by Kanchanavatee \textit{et
al.} in 2014 \cite{Kanchanavatee2014}. These studies show that doping is
possible up to $x = 1.2$ with no change in the structure and only a small
increase in the lattice constant compared to the large decrease seen for Fe
doping. Accompanying this small expansion of the lattice, the transition
temperature dramatically increases up to a maximum of 50~K by $x = 1.2$. From
resistivity and specific heat measurements Kanchanavatee \textit{et al.}
hypothesize a transition out of the HO state at $x = 0.2$. However, this study
again did not involve any microscopic probes of magnetism and hence the true
evolution of the magnetic ground state of URu$_{2-x}$Os$_x$Si$_2$ is still an
open question.

In this paper we present the results of $\mu$SR and susceptibility measurements
on URu$_{2-x}$T$_x$Si$_2$ (T = Fe, Os) single crystals for doping levels up to
$x = 1$. Our measurements demonstrate that an antiferromagnetic state arises for
both of these compounds at low doping levels and highlight the importance of
hybridization to fully understand the evolution of magnetic order in this
system.

\section{Experimental Methods}
Samples measured in this study were single crystals grown by the Czochralski
method at McMaster University from starting materials of depleted U,
Ru(99.95\%), Fe(99.99\%), Os(99.8\%), and Si(99.9999\%). These growths were
performed in a tri-arc furnace from a water-cooled copper hearth under Argon
gettered at 900$^o$C. After the growths, crystallinity was confirmed and sample
alignment performed by Laue x-ray scattering measurements.

Magnetic susceptibility measurements were performed on cleaved plates of the
crystals in a Quantum Design MPMS XL-3. These measurements provide a measure of
the transition temperature from the paramagnetic state to hidden order or
antiferromagnetism, however they cannot readily distinguish hidden order from
antiferromagnetism.

$\mu$SR is a sensitive microscopic magnetic probe that can distinguish
antiferromagnetism from hidden order, but cannot readily distinguish hidden
order from paramagnetism. In this technique, spin polarized positive muons are
injected one at a time into a sample where they penetrate a few hundred $\mu$m,
rapidly thermalize, and stop at a Coulomb potential minima in the material. Once
stopped, each muon spin precesses in the local magnetic field until it decays
with average lifetime of 2.2~$\mu$s and emits a positron preferentially in the
direction of the muon spin at the time of decay. Detectors on either side of the
sample register the decay of the positron and record the time interval between
muon injection and decay. From many such events, a histogram of positron counts
in both detectors, $N_R$ and $N_L$, as a function of muon decay time is
generated. Using these two histograms the asymmetry, $A$, is defined as $A =
\frac{N_R - N_L}{N_R + N_L}$. This quantity gives a measure of how the muon
polarization changes over time, and is limited by the physics of muon decay and
instrumental factors to a maximum of about 0.3. The true maximum in any given
experiment is determined from the total oscillating asymmetry seen after
applying a small field transverse to the muon polarization in the paramagnetic
regime.

In zero applied magnetic field, paramagnetic samples, where there is no static
magnetism and the spin dynamics are very fast, will show a nearly
time-independent asymmetry, with deviations from this caused by nuclear moments.
The HO state will also have this signature, as there is no ordering of magnetic
moments to produce a local magnetic field. By contrast, long-range ordered
magnetic states such as antiferromagnetism will show an oscillating asymmetry
where the frequency gives the strength of the internal field at the muon
stopping site, provided this field is not parallel to the initial muon
polarization. The ratio of the maximum amplitude of the oscillating asymmetry to
the instrumental maximum gives the fraction of the sample that is in the
magnetic state (the magnetic volume fraction). The amplitude of this oscillating
signal damps down over time as a result of inhomogeneities and dynamics of the
internal field.

Our $\mu$SR measurements were performed on the M15 and M20 beam lines at TRIUMF
laboratory in Vancouver. The LAMPF time-differential spectrometer was used,
which provides a He-4 cryostat for temperatures between 2 and 300K and a time
resolution of 0.2~ns in an ultra low background apparatus. This apparatus vetoes
muons that miss the sample, ensuring that almost all of the measured positrons
come from muons that stop in the sample. For these measurements the single
crystals were cleaved into slices roughly 0.5mm thick along the c-axis which
were then mounted in a mosaic covering 1-2~cm$^2$ on thin mylar tape. The c-axes
were coalligned facing the muon beam but no attempt was made to coalign the
samples in the a-b plane. We fit our $\mu$SR data using the $\mu$SRfit software
package \cite{musrfit}.

\section{Results}

\begin{figure}[t]
\includegraphics[width=\columnwidth]{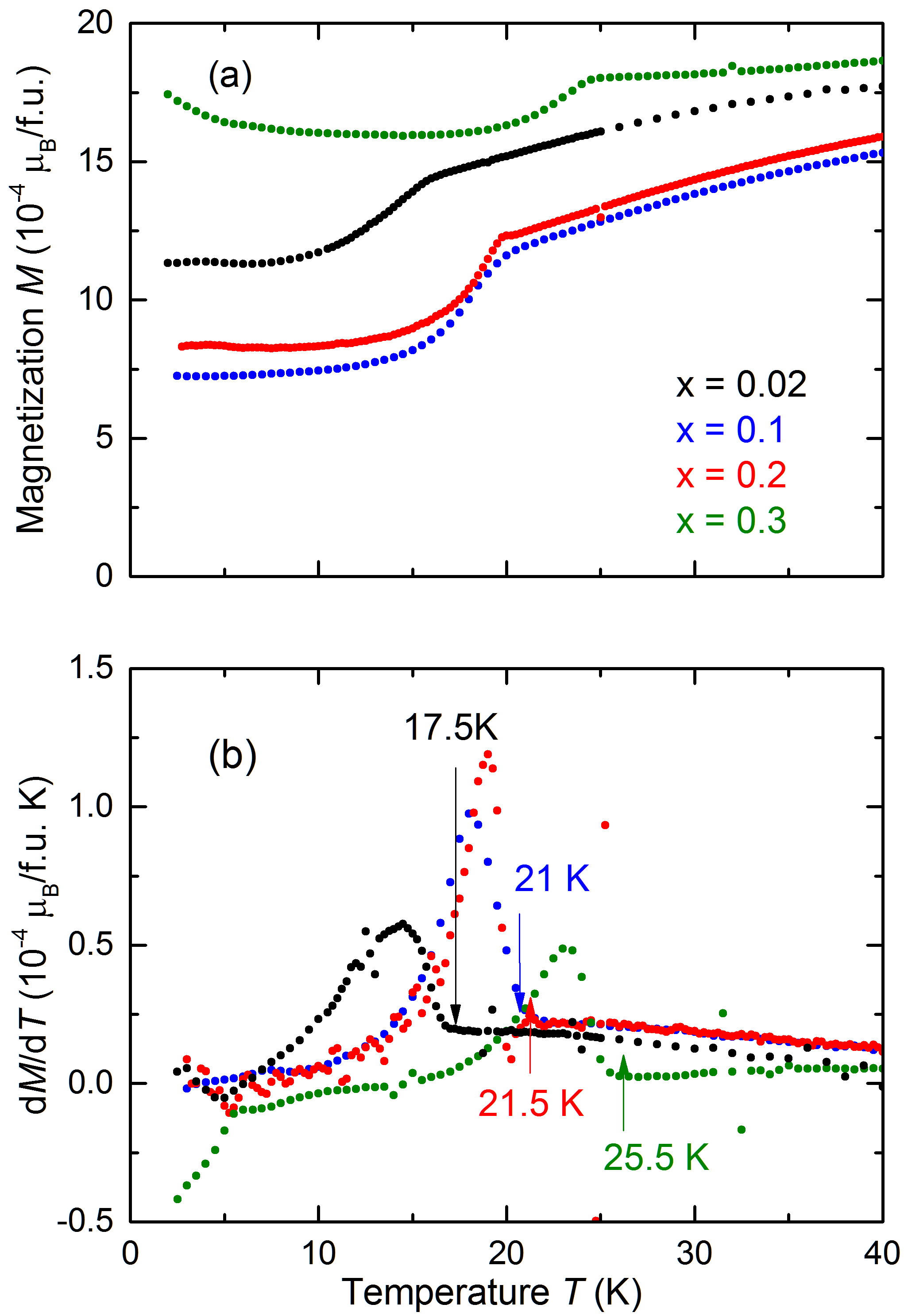}
\caption{(a) URu$_{2-x}$Fe$_x$Si$_2$ magnetization data measured in a field of H
= 0.1~T $\parallel$~$c$. (b) Temperature derivative of the data shown in (a)
Arrows on plot (b) show the measured transition temperatures.}
\label{fig:FeSQUID}
\end{figure}

\begin{figure}[t]
\includegraphics[width=\columnwidth]{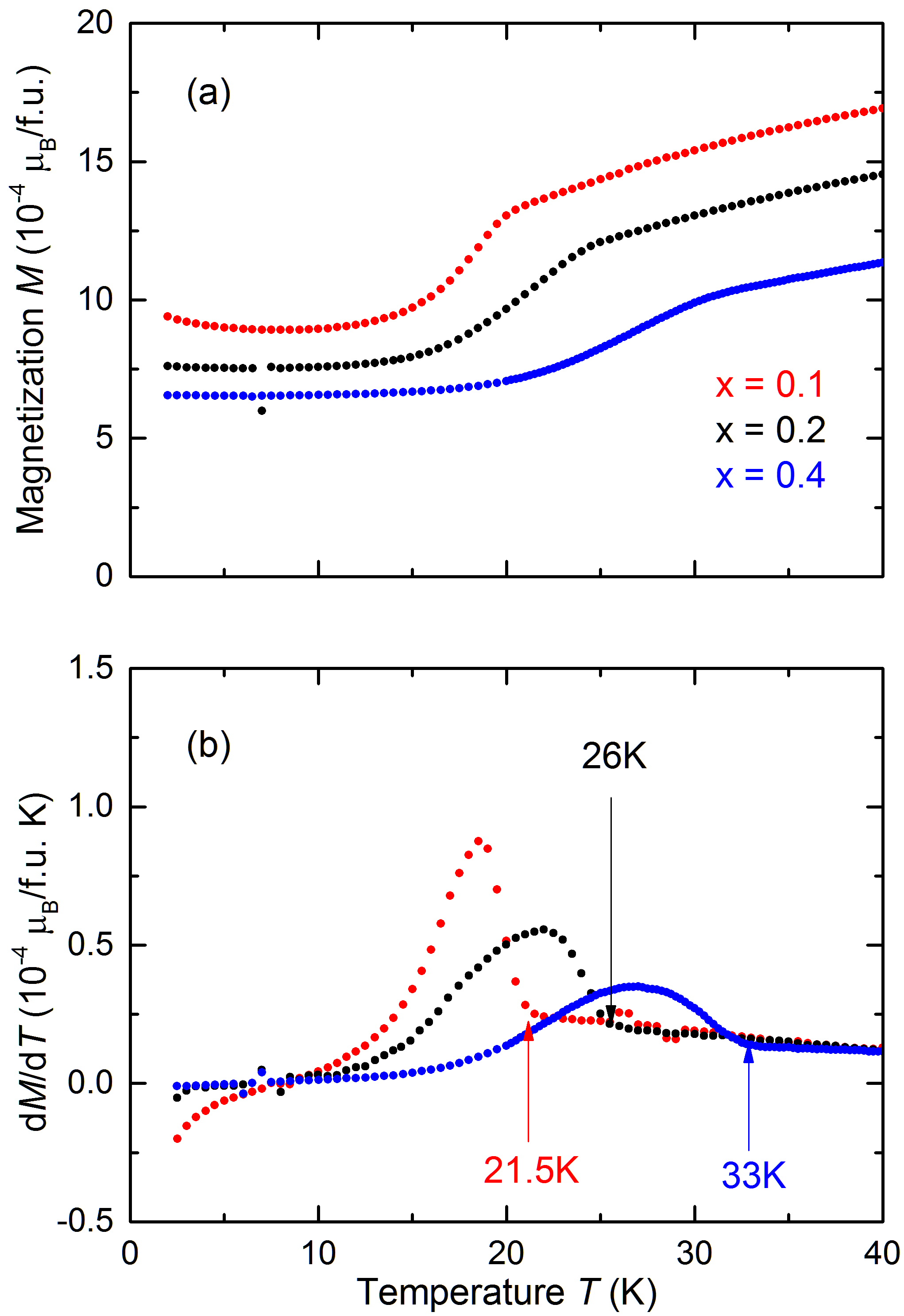}
\caption{(a) URu$_{2-x}$Os$_x$Si$_2$ magnetization data measured in a field of H
= 0.1~T $\parallel$~$c$. (b) Temperature derivative of the data shown in (a).
Arrows on plot (b) show the measured transition temperatures.}
\label{fig:OsSQUID}
\end{figure}

Figures \ref{fig:FeSQUID} and \ref{fig:OsSQUID} show the results of the
magnetization measurements in a field of 0.1~T ($H$~$\parallel$~ $c$) on the Fe
and Os doped samples respectively. No significant differences were observed in
any of these samples between field-cooled and zero field-cooled and hence only
one set of measurements are shown. In these measurements a kink in the
susceptibility indicates the transition into either the HO or LMAF states. The
lower panels of these figures show plots of d$M$/d$T$ to allow a more accurate
determination of the temperature of this kink.

The measurements on the Fe doped samples show little change in the character of
the transition with doping; the transition remains a relatively sharp peak in
d$M$/d$T$ up to higher dopings. The sharp peak is consistent with measurements
on polycrystalline samples presented by Kanchanavatee \textit{et al.}
\cite{Kanchanavatee2011}. However, our measurements on single-crystal samples do
not show evidence of the significant second peak seen in some of the
polycrystalline samples in ref. \cite{Kanchanavatee2011}. This likely indicates
that those features were spurious results arising from disorder in the
polycrystalline samples, as was also proposed by Das \textit{et al.}
\cite{Das2015}. Additionally, our x = 0.3 sample shows a significant low
temperature upturn in the magnetization as well as the highest overall
magnetization. During the crystal growth of this sample, a small number of
needle-like protrusions were noticed on the outside of the crystal, likely
indicating some phase separation that would cause a paramagnetic background in
the magnetization measurements, as observed. We attribute this to a lower than
nominal silicon level in the melt arising from evaporation as silicon has the
highest vapor pressure of the elements present \cite{Honig1962} and the growth
for this doping was held at high temperature for a significantly longer period
than the others.

The measurements on our Os doped samples show a somewhat different evolution in
the character of the transition with doping. Rather then staying as a sharp
peak, the transition broadens significantly and shifts to higher temperature as
the doping level increases. This is consistent with the broadened transition
seen in polycrystalline samples at x = 0.3 and 0.4 \cite{Kanchanavatee2014}.
\begin{figure}[t]
\includegraphics[width=\columnwidth]{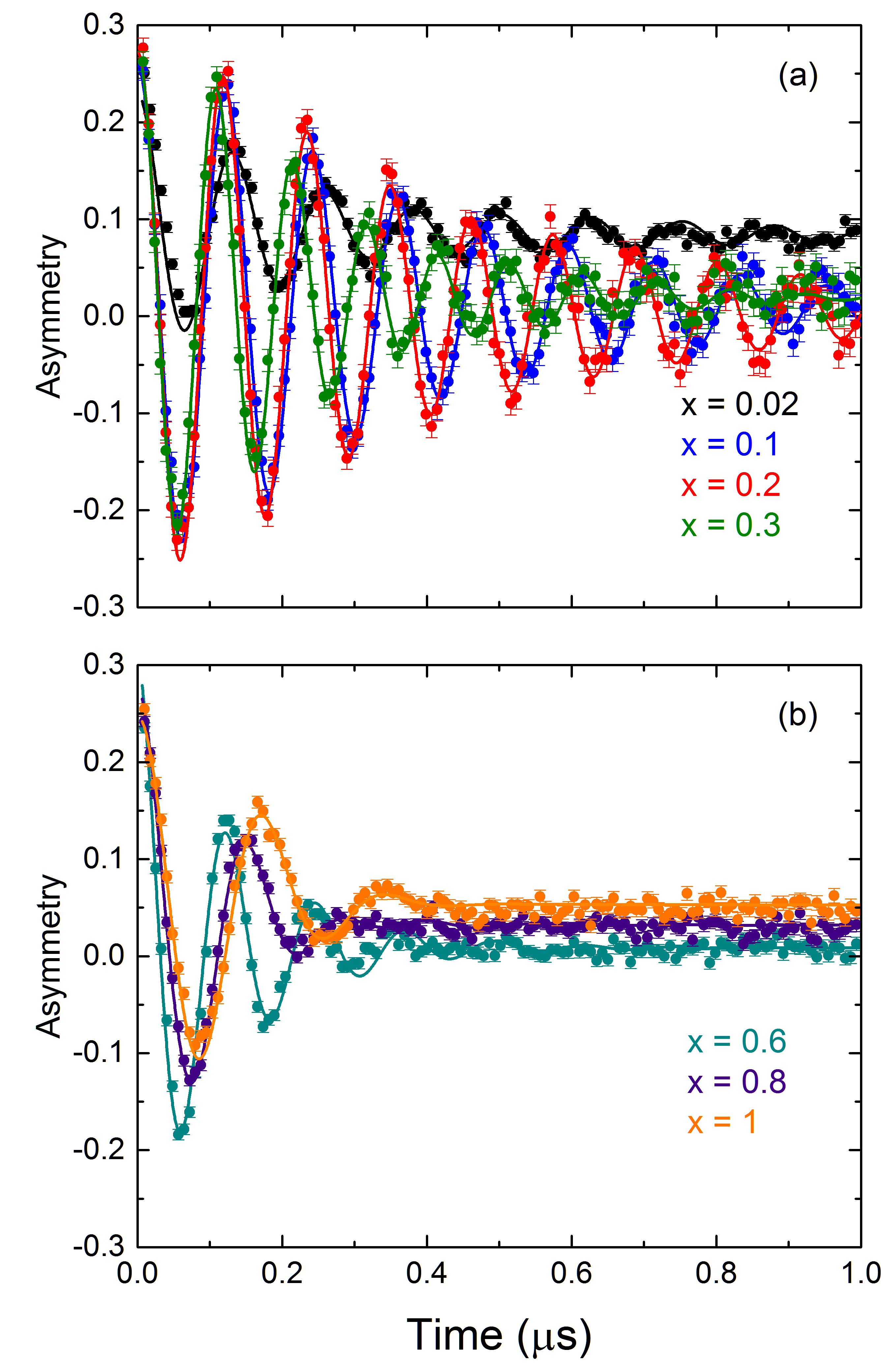}
\caption {URu$_{2-x}$Fe$_x$Si$_2$ $\mu$SR Data measured at $T = 2$~K in zero
applied external field with the muon spins initially perpendicular to the
c-axis. Solid lines in (a) show fits to Eq. \ref{eq:musr2} and those in (b) show
fits to Eq. \ref{eq:musr1}}
\label{fig:Femusrdat}
\end{figure}

$\mu$SR data for the Fe samples at 2~K measured with the muon spins initially
perpendicular to the c-axis of the crystals in zero applied field (ZF) is shown
in Fig. \ref{fig:Femusrdat}. Measurements in panel (b) were taken with higher
statistics to better resolve the faster relaxing signal. This data exhibits
clear oscillations for all samples, indicating that there is static magnetism
with the field along the c-axis at the muon stopping site. The amplitude of the
oscillations for the x = 0.02 sample is significantly lower than for the others
and the asymmetry is shifted upwards by a non-relaxing component. This indicates
that the magnetic volume fraction is lower in this sample.

We found that applying a small field parallel to the c-axis to any of these
samples splits the observed internal field into two components separated by
twice the applied field. This indicates that the magnetic order in these samples
is antiferromagnetic. We also performed some measurements with the muon spins
parallel to the c-axis that show no oscillations for the low doping samples,
indicating that the internal field is only along the c-axis within the accuracy
of our alignment. This is consistent with the antiferromagnetic phase seen in
URu$_2$Si$_2$ under hydrostatic pressure \cite{Amitsuka1999} and by Das
\textit{et al.} in neutron scattering measurements on URu$_{2-x}$Fe$_{x}$Si$_2$
\cite{Das2015} which has magnetic moments along the c-axis. However, it should
be noted that while the direction of the internal field often matches the moment
direction, this is not always the case and full comparison depends on knowledge
of the muon stopping site which we do not have.

Despite the apparent similarity of this antiferromagnetic state to that of
URu$_2$Si$_2$ under hydrostatic pressure, we found that the fitting of the ZF
data at low doping was significantly improved with a two component fit compared
to the single component fit used by Amato \textit{et al.} for the pure compound
\cite{Amato2004}. We therefore fit the data for x = 0.02-0.3 shown in Fig.
\ref{fig:Femusrdat} (a) using the equation,

\begin{equation}
\begin{split}
A = & A_T \left[ 0.5F \left(\cos(2\pi\gamma_{\mu}B t)e^{-0.5(\sigma_1
t)^2}\right. \right. \\
& \left. \left.+ \cos(2\pi\gamma_{\mu}B R t)e^{-0.5(\sigma_2 t)^2}\right) +
(1-F)\right].
\end{split}
\label{eq:musr2}
\end{equation}

In this model the ratio of the asymmetries of the two components was fixed to
0.5 for simplicity as fits with free asymmetry were found to refine to values
near to 0.5. Addition of a second frequency for the higher dopings x = 0.6 to
1.0 did not improve the fits compared to the single component model given by the
equation,

\begin{equation}
A = A_T \left( F \cos(\gamma_{\mu}B  t)e^{-0.5(\sigma t)^2} + (1-F)\right).
\label{eq:musr1}
\end{equation}

Therefore, Eq. \ref{eq:musr1} was used to fit the data in Fig.
\ref{fig:Femusrdat} (b). In these equations $A_T$ is the total asymmetry, $B$ is
the larger internal field, $\gamma_{\mu}$ = 135.538 MHz/T is the muon
gyro-magnetic ratio, $R$ is the ratio between the internal fields at the two
muon sites, $F$ is the magnetic volume fraction, and the $\sigma_i$ are the
relaxation rates. For each of the fits $A_T$ and $R$ were temperature
independent parameters for each sample and the other parameters were allowed to
vary with temperature.

The field ratio, $R$, varies between samples with no obvious doping dependence
as shown in Table \ref{tab:Fefit}. However, the relaxation rate also varies
erratically from sample to sample, likely from differing amounts of disorder,
and this will affect the fitting of a second frequency. Table \ref{tab:Fefit}
also shows the substantially larger single relaxation for the higher doped
samples which obscures any possibility of fitting a second field to these data.
We expect that a second frequency may still be present but increased disorder
from growing crystals at high doping levels makes it impossible to distinguish.

\begin{table}

  \begin{tabular}{| c | c | c | c |}
      \hline
Doping $x$ & $R$ & $\sigma_1$ ($\mu$s$^{-1})$ at 2K & $\sigma_2$ ($\mu$s$^{-1})$
at 2K \\
\hline
0.02 & 0.86 $\pm 0.01$ & 2.39 $\pm 0.08$ & 5.37 $ \pm 0.29$   \\
0.1 & 0.942 $\pm 0.008$  & 1.95 $ \pm 0.04$ & 4.98 $\pm 0.13$  \\
0.2 & 0.93 $\pm 0.02$  & 1.763 $\pm 0.04$ & 4.39 $\pm 0.15$ \\
0.3 & 0.891 $\pm 0.004$ & 2.822 $\pm 0.07$ & 4.51 $\pm 0.12$  \\
0.6 & - & 7.57 $\pm 0.1$ & -   \\
0.8 & - & 9.33 $\pm 0.18$ & -   \\
1 & - & 7.04 $\pm 0.1$ & -   \\
\hline
  \end{tabular}
\caption{Relaxation rates used to fit the 2K $\mu$SR data on Fig.
\ref{fig:Femusrdat} for URu$_{2-x}$Fe$_x$Si$_2$ and the temperature independent
ratio of the two internal fields used in fits to Eq. \ref{eq:musr2}}
\label{tab:Fefit}
\end{table}

\begin{figure}[ht]
\includegraphics[width=\columnwidth]{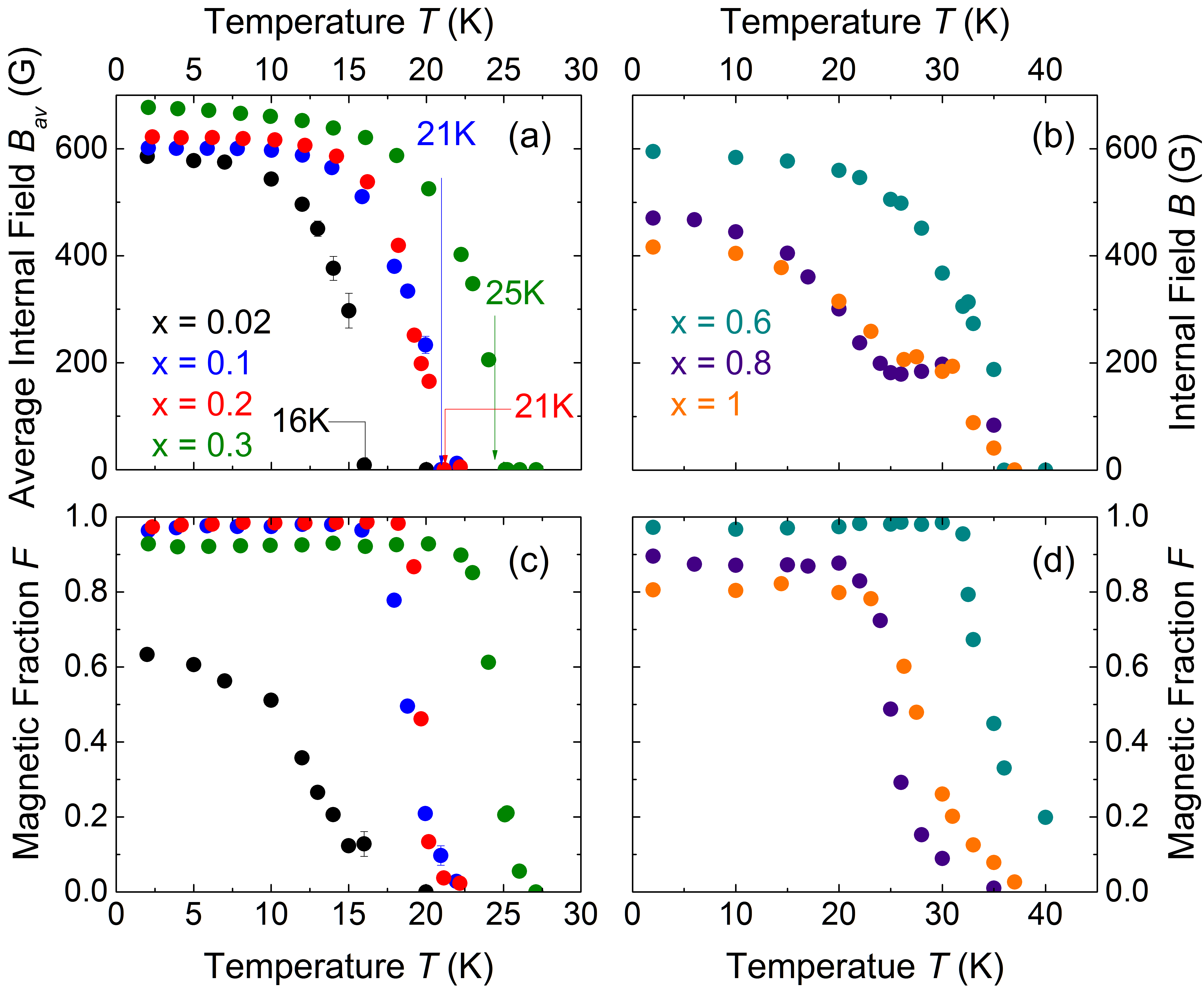}
\caption{Fitting parameters for the $\mu$SR data of URu$_{2-x}$Fe$_x$Si$_2$
measured in zero applied field with the muon spins perpendicular to the c-axis.
(a) Average internal field $B_{av} = 0.5(B + RB)$ for dopings x = 0.02 (black),
0.1 (blue) 0.2 (red) and 0.3 (green). (b) Internal field $B$ for dopings x = 0.6
(yellow), 0.8 (purple) and 1 (orange). (c) Magnetic volume fraction for dopings
x = 0.02 (black), 0.1 (blue) 0.2 (red) and 0.3 (green). (d) Magnetic volume
fraction for dopings x = 0.6 (yellow), 0.8 (purple) and 1 (orange). }
\label{fig:Feparam}
\end{figure}

Figure \ref{fig:Feparam} shows plots of the fit average internal field ($0.5(B +
RB)$ for the lower dopings) and magnetic fraction $F$. In all samples the
internal field smoothly decreases from a maximum at low temperature to zero at
the transition, showing second order behavior. The magnetic fraction for all
samples except for the x = 0.02 is mostly temperature independent up until the
transition where a sharp fall off occurs. This fraction is close to 1 for the x
= 0.1 - 0.3 samples and slightly lower for the higher dopings. In contrast to
the others, the x = 0.02 sample shows a substantially reduced $F$ of 0.63 at
2~K. Furthermore, this sample shows different temperature dependence with a
smooth continuous drop off in the magnetic fraction over the entire temperature
range. This may indicate a continuous volume-wise transition out of the AF state
as a function of temperature.

\begin{figure}
\includegraphics[width=\columnwidth]{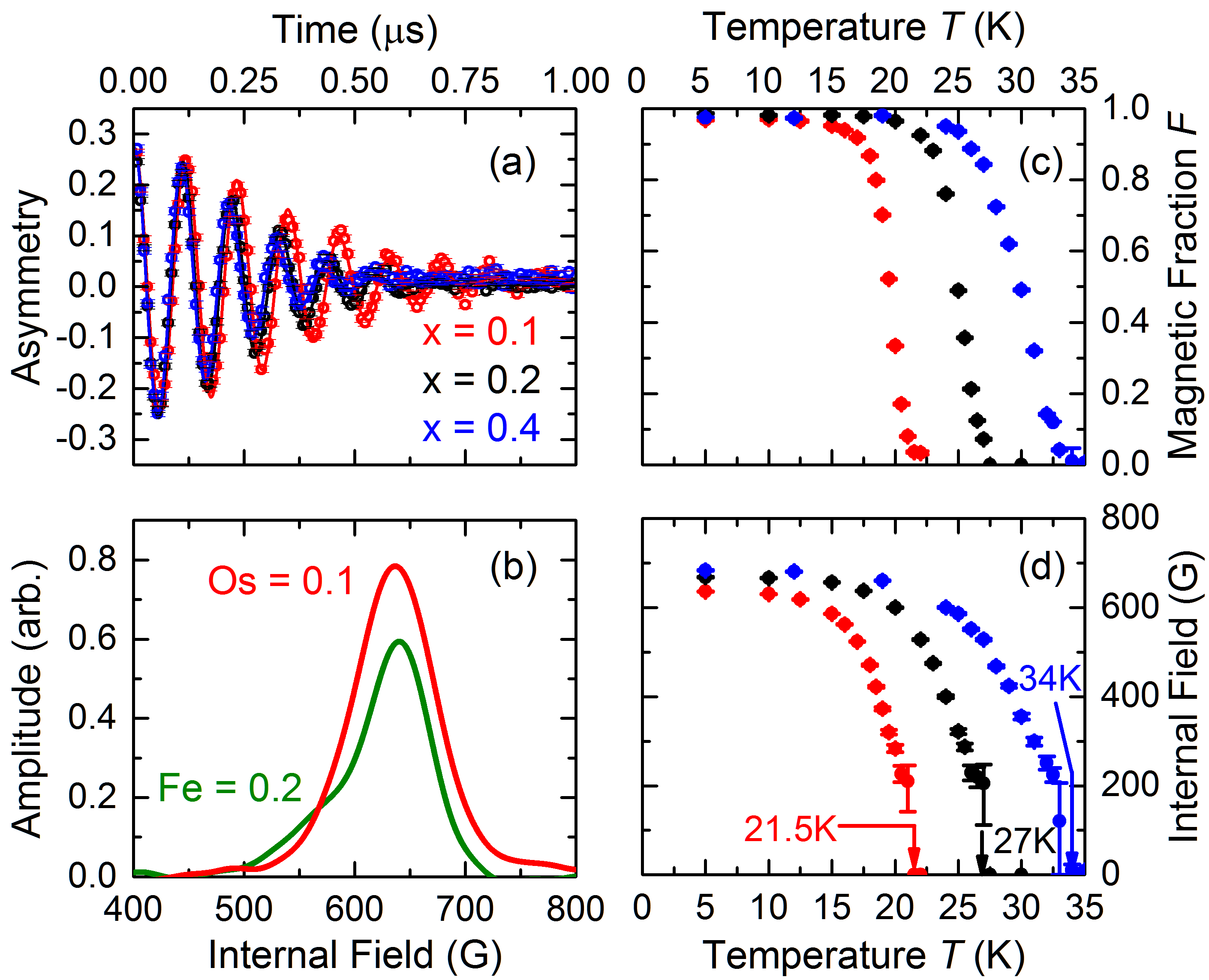}
\caption{URu$_{2x-}$Os$_x$Si$_2$ $\mu$SR data and fitting in zero external field
measured with muon spins initially perpendicular to the c-axis for x = 0.1
(red), 0.2 (black) and 0.3 (blue). (a) $\mu$SR data measured at 5~K, (b) Fourier
transform of URu$_{1.9}$Os$_{0.1}$Si$_2$ (red line) data measured at 5K and
URu$_{1.8}$Fe$_{0.1}$Si$_2$ (green line) data measured at 2K, (c) magnetic
volume fraction $F$ (d) internal field $B$. Solid lines in (d) correspond to
fits to Eq. \ref{eq:musr1}.}
\label{fig:Osmusr}
\end{figure}

$\mu$SR data collected at $T = 5$~K in zero field with the muon spins initially
perpendicular to the c-axis for the Os doped samples are shown in Fig.
\ref{fig:Osmusr} (a). For these samples the data again shows clear oscillations
indicating similar static order. However, there is no evidence for a second
internal field component in these samples. Therefore, we fit the data using Eq.
\ref{eq:musr1} and show the internal field and magnetic volume fraction in Fig.
\ref{fig:Osmusr} (c) and (d). These plots show similar temperature dependence to
the Fe doped samples again indicating a second order transition in all samples.

The comparison of two internal fields for Fe at low doping compared to one
frequency in Os is illustrated by the Fourier transform in Fig. \ref{fig:Osmusr}
(b). This plot shows that while two frequencies appear in the Fe sample, the
overall linewidth is similar for the Os sample. This means that the appearance
of a second field for Os samples could be masked by the larger linewidth.
Similarly, Table \ref{tab:Fefit} shows that the relaxation rate (linewidth) is
much higher in the heavily doped Fe samples where two frequencies are not
resolved. This is likely a result of chemical disorder in the samples and would
explain why we cannot see two frequencies in these cases. A similar mechanism
may explain the lack of a second field for the measurements under pressure done
by Amato et al \cite{Amato2004}. In this case, the pressure was applied with an
anvil cell using a transmitting medium that would be frozen at the relevant
temperatures. This can cause non-uniformities in the applied pressure
\cite{Yu2009}, which would introduce inhomogeneity in the samples, increasing
the linewidth and masking the appearance of a second frequency. Furthermore, in
any experiment with a pressure cell many muons are stopped outside the sample.
This drops the signal to noise ratio of the data, further reducing the ability
to resolve a possible second frequency. These explanations would allow for the
magnetic state to be nearly identical in our Fe and Os samples as well as the
pure URu$_2$Si$_2$ measured under pressure, despite the apparent differences in
fitting.

The presence of a second internal field in any of these measurements indicates
that the muons stop at two magnetically distinct sites at equivalent or
near-equivalent Coulomb potential minima. The second magnetic site could either
be explained by a more complex magnetic structure that breaks one of the
symmetries of the underlying crystal lattice, or structural effects creating two
muon sites. If this does appear only for doping, one possibility is that the
Fe/Os atoms are being magnetically polarized and contributing to the moment seen
by the muons. However, our measurements indicate that the relative volume
fraction of the two magnetic sites is close to 50/50, which would not be
expected if one of these was coming from the 1-15\% doping. Furthermore,
UFe$_2$Si$_2$ and UOs$_2$Si$_2$ are both non-magnetic so we would not expect Fe
and Os polarization \cite {Szytula1988, Palstra1986}. Future detailed
measurements of the temperature and doping dependence of the lattice parameters
and structure symmetry would help clarify this issue.

\section{Discussion}

The fit parameters in Figure \ref{fig:Feparam} and \ref{fig:Osmusr} show two
important features. First, for most samples the low temperature magnetic volume
fraction is close to one. This tells us that the magnetism we see must be
attributed in each case to the bulk of the sample rather than a small impurity
effect. The small non-magnetic volume that does appear could be attributed to
muons stopping in parts of the sample holder rather than the sample itself or
slight mis-alignment of the samples with respect to the incoming muon beam. In
the heavily doped samples where the volume fraction appears somewhat reduced, a
small signal also appears in measurements with the muon spin rotated parallel to
the aligned c-axis. Misalignment would explain both the signal in the $\mu || c$
measurements and the reduced signal / volume fraction for $\mu$ $\perp$ $c$ as
the measured asymmetry varies as $\sin\theta$ where $\theta$ is the angle
between the muon spins and the internal field.

\begin{figure}
\includegraphics[width=\columnwidth]{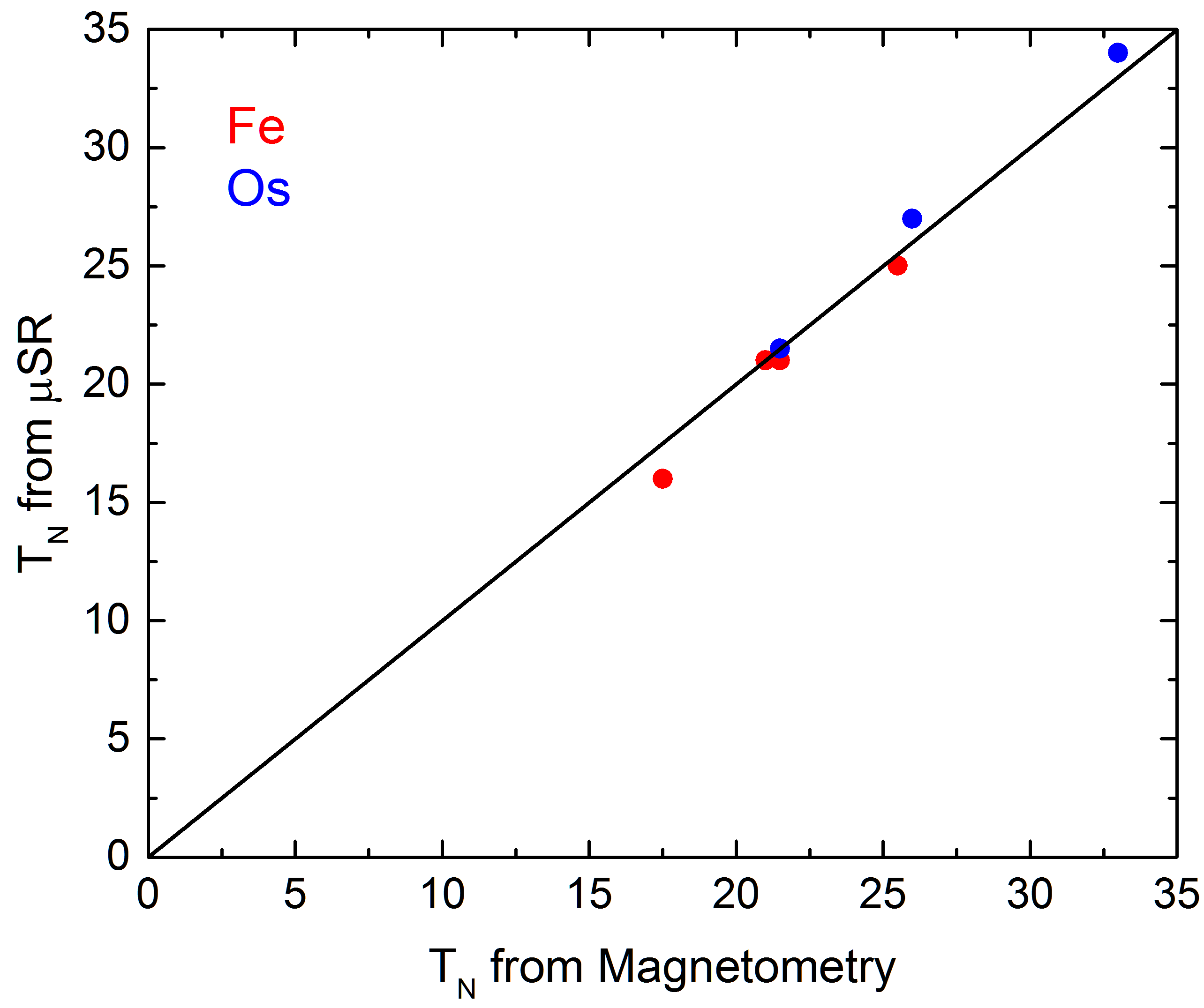}
\caption{Comparison of the transition temperatures measured by susceptibility to
those measured by $\mu$SR for Fe doped samples (red) and Os doped samples
(blue). The solid black line shows the expected 1:1 correspondence.}
\label{fig:TvT}
\end{figure}

Second, with the exception of the x = 0.02 Fe doped sample, the internal field
falls off smoothly as a function of temperature to zero at a transition
temperature consistent with that shown by the magnetization measurements. This
indicates that the system transitions directly from the magnetically ordered to
paramagnetic (PM) states without the transition through HO that has been seen
for intermediate pressures applied to URu$_2$Si$_2$ \cite{Amato2004,
Hassinger2008, Bourdarot2014, Butch2010}. In the Fe = 0.02 sample the transition
temperature from $\mu$SR is 1.5~K lower than that measured by SQUID. This small
discrepancy is unlikely to be caused by thermometry differences, as the same
thermometry was used for $\mu$SR measurements of all other samples where the
transition temperatures appear more consistent as shown by Fig \ref{fig:TvT}.
Furthermore, the distinctly different temperature dependence in the magnetic
volume fraction of this sample compared to the others leads us to believe that
the magnetic state may not the same. One explanation for these discrepancies is
if this sample is in a mixed HO/AF state below 17.5~K, with the volume fraction
of the AF state decreasing up until 16~K leaving a pure HO state in a 1.5~K
range between 16 and 17.5~K. In the pressure-temperature phase diagram of pure
URu$_2$Si$_2$ there exists a small temperature range where decreasing
temperature first causes a transition into hidden order and then to
antiferromagnetism, so it would not be unexpected to find a similar region at
low Fe dopings in our system. However, as the transitions measured by both
techniques are reasonably broad, and the temperature discrepancy is small, it is
not possible to draw firm conclusions about the existence of both HO and AF at
different temperatures in this sample. Further measurements on this doping with
other techniques, particularly those that show a direct signature of the HO
state such as inelastic neutron scattering, which has been used to distinguish
the two under pressure \cite{Butch2010}, will be required to clarify this issue.

\begin{figure}
\includegraphics[width=\columnwidth]{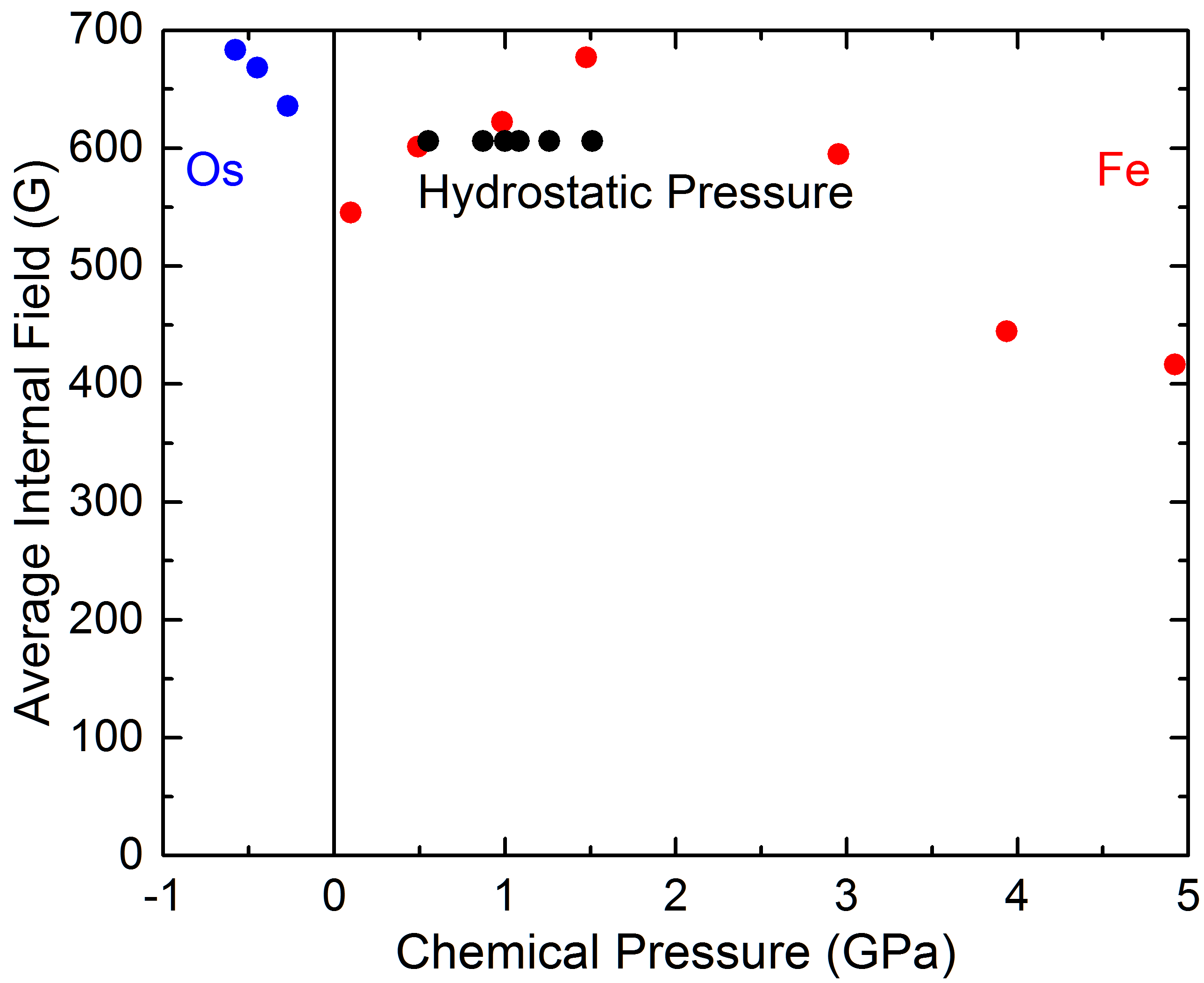}
\caption{Measured internal fields as a function of effective chemical pressure
for Fe doped (red) and Os doped (blue), and as a function of applied hydrostatic
pressure for pure URu$_2$Si$_2$ (black) from Ref. \cite{Amato2004}.}
\label{fig:BvX}
\end{figure}

The overall behavior of the $\mu$SR data presented in this work is similar to
that seen in measurements on URu$_2$Si$_2$ under hydrostatic pressure
\cite{Amato2004}. However, there are some notable differences. First, while the
internal field measured at low temperature is comparable to that of Amato
\textit{et al.}, our measured internal fields for both Os and Fe increase with
doping, while the internal field above some critical pressure is constant for
URu$_2$Si$_2$ under pressure \cite{Amato2004}. This difference in behavior is
clearly demonstrated in Fig. \ref{fig:BvX} showing the low temperature internal
fields measured for all samples in this study plotted as a function of chemical
pressure along with the data from Amato \textit{et al.} For this plot the
effective chemical pressure, $P_{ch}$, was calculated using $P_{ch} =(\Delta
V)/(V_0)/\kappa$, where $\kappa = 5.2 \times 10^{-3} $ GPa$^{-1}$ is the bulk
modulus for pure URu$_2$Si$_2$ \cite{Kuwahara2003}, $\Delta V$ is the unit cell
volume change from pure URu$_2$Si$_2$ taken from the crystallographic data in
Refs \cite{Kanchanavatee2011, Kanchanavatee2014} using our nominal doping
levels, and $V_0$ is the unit cell volume of pure URu$_2$Si$_2$. This figure
also indicates that the appearance of magnetic order cannot be attributed to
chemical pressure across this system as the Os doped samples show similar
internal fields at effective chemical pressures that are negative and whose
magnitude is significantly lower than that for Fe doping. The Fe $x = 0.02$
sample also still shows magnetic order despite being at an effective chemical
pressure less than a quarter of the pressure required to generate the LMAF in
pure URu$_2$Si$_2$.

It has been proposed in the past that the transition between HO and LMAF is
governed by the $\eta = c/a$ ratio as has been demonstrated for superconducting
transitions in other f electron compounds \cite{Pfleiderer2009}, rather than
uniform shrinking of the unit cell \cite{Yokoyama2005, Kanchanavatee2014}. While
both Fe and Os doping do increase $\eta$, the change is an order of magnitude
smaller for Os doping than is seen for Fe doping or applied pressure. This
indicates that the change in $\eta$ alone cannot explain the development of
magnetic order.

\begin{figure}
\includegraphics[width=\columnwidth]{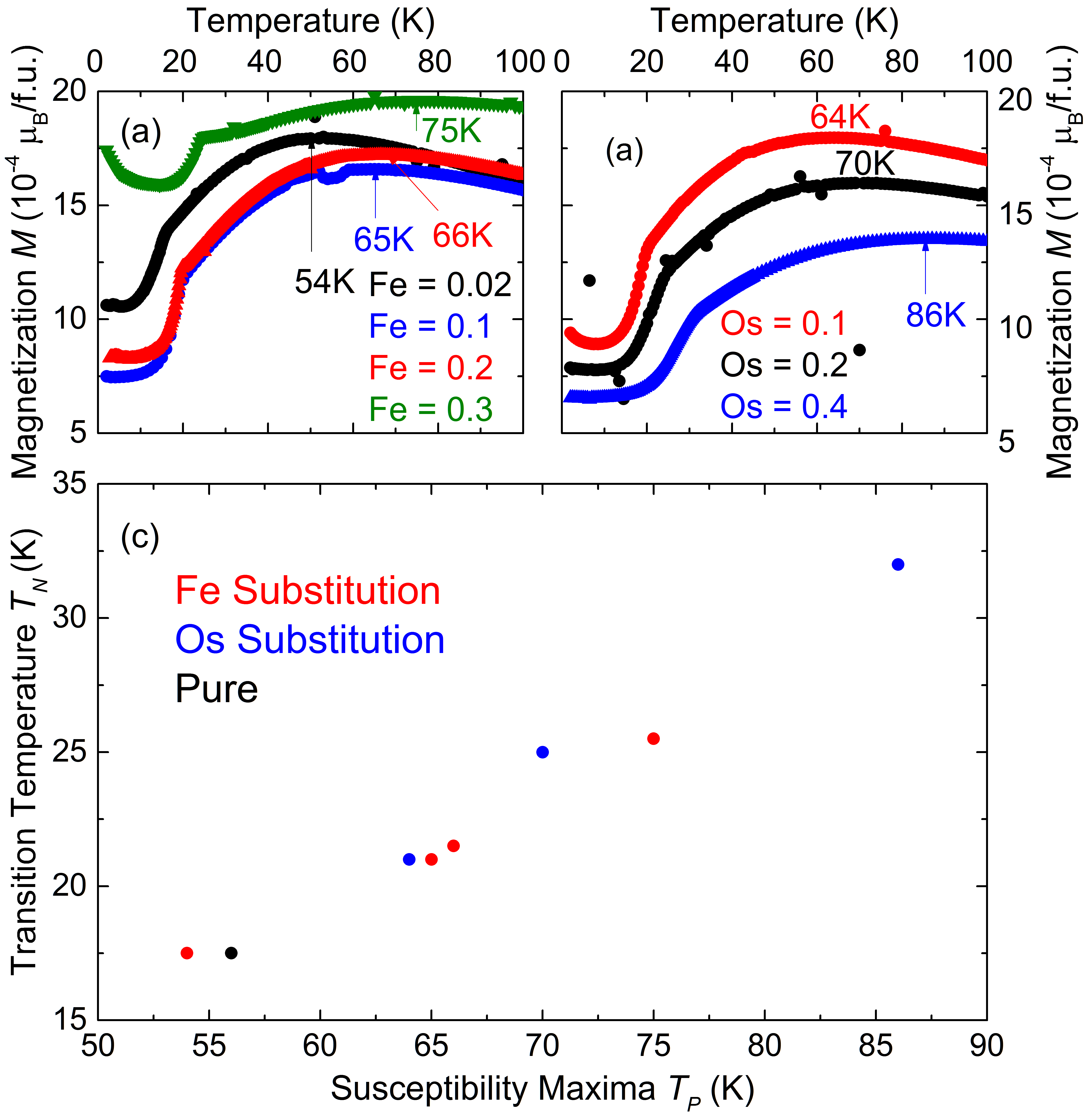}
\caption{High temperature susceptibility data showing the broad maxima that
appears for (a) Fe doped samples and (b) Os doped samples. Panel (c) shows a
plot of $T_N$ from susceptibility vs. the temperature of this susceptibility
maxima for Fe doped (red) and Os doped (blue).}
\label{fig:TnTp}
\end{figure}

Susceptibility data on the lower doped samples show a clear broad maximum at
high temperatures, shown in Fig \ref{fig:TnTp} (a) and (b). Such a maximum is
expected for heavy fermion compounds and arises from the crossover from
local-moment magnetism at high temperature to the heavy fermion state at low
temperatures caused by the hybridization of the conduction electrons with the
core $f$-electrons \cite{Stewart1984}. Hence, the temperature of this crossover,
$T_{max}$, can be taken as a rough proxy for the strength of hybridization in
these systems.
Our data shows an increase in $T_{max}$ with doping for both Os and Fe, which
suggests that hybridization between the U $f$-electrons and the valence
electrons increases with doping for both cases. Furthermore, Fig. \ref{fig:TnTp}
shows a similar linear correlation between $T_N$ and $T_{max}$ in both cases.
This points to hybridization as the driving force behind the evolution of
magnetic order in these systems. In contrast, measurements by others of
$T_{max}$ as a function of pressure for pure URu$_2$Si$_2$ show a
pressure-independent value of approximately 60K over a range where the magnetic
transition temperature increases from 16 to 18.5K \cite{Nishioka2000}. This
difference further emphasizes that the evolution of magnetism in our systems is
not driven by chemical pressure alone.

Our results for Fe doping show some discrepancies with those reported recently
by Das \textit{et al.} using neutron scattering on crystals that should be
similar to ours \cite{Das2015}. First, our internal fields increase with doping,
while the results of Das \textit{et al.} show either doping independence or a
slight decrease with doping. Second, our measured internal field is roughly
consistent with URu$_2$Si$_2$ under pressure, while Das \textit{et al.} report a
magnetic moment up to twice that measured for the LMAF in URu$_2$Si$_2$.
Finally, we see similar magnetism down to low doping levels while Das \textit{et
al.} see weakening of the magnetism below x = 0.1.

The first discrepancy could be explained by slight changes to the muon stopping
site with doping. If the muons systematically stop closer to the magnetic U
atoms as the Fe doping increases, this would cause a small increase in our
observed internal field even if the magnetic moments are constant or slightly
decreasing. However, in a simplistic viewpoint the dopant Fe atoms should have
electron orbitals with smaller spatial extent than the Ru, and hence one would
expect the muon stopping sites to move closer to the Fe atoms and further from
the magnetic U ions. This would cause a decrease in the measured internal field
rather than an increase. Detailed numerical calculations of the likely muon
stopping sites would be required to quantitatively determine the effect of the
Fe doping. Another explanation for the doping dependence is Fe site magnetism
contributing to the internal field, which could potentially be clarified with
M\"{o}ssbauer measurements that could directly measure the Fe magnetism.

The second discrepancy is difficult to reconcile. While $\mu$SR cannot provide a
numerical value of the magnetic moment without knowledge of the muon stopping
site which we do not have, the comparison between the measured internal fields
of samples with very similar structures should give a good idea of how the
magnetic moment changes between these samples. Therefore, the Fe = 0.1 sample
should be reasonably comparable to the pure compound under pressure and hence
seeing a similar internal field here should indicate that the magnetic moments
are the same. While the doping could change the muon stopping site somewhat
between pure URu$_2$Si$_2$ and the Fe = 0.1 sample, the structure and lattice
constants remain mostly the same and it seems unlikely that this would be a
large enough effect to cut the measured internal field in half to make our
results consistent with the magnetic moment measured by Das \textit{et al.} One
possibility is that there is signal intensity at the magnetic Bragg peak
positions from multiple scattering that Das \textit{et al.} may not have taken
into account and would artificially inflate the calculated magnetic moments.
Recent neutron diffraction on a number of the samples from this work utilize a
different method of normalizing the data that reduces the effect of multiple
scattering \cite{Williams2016}. These results find moments that are more
consistent with the values obtained under pressure, possibly showing that
incorrect normalization is the cause of our discrepancy with the work of Das
\textit{et al.}

The final discrepancy of our data showing magnetism down to lower doping levels
may come down to slight variations in doping levels or internal strain between
different crystals; however, the results are not entirely inconsistent. Das
\textit{et al.} report that there is some magnetic scattering still appearing in
the lower doped samples, it is just substantially reduced. This could come from
magnetic moments that are the same as those measured in higher doping samples,
but with a reduced magnetic volume fraction, as the Bragg peak intensity cannot
distinguish volume fraction from magnetic moment. A reduced volume fraction with
similar magnetic moment would be qualitatively consistent with the results we
show for our nominal Fe = 0.02 sample.

\section{Conclusion}

In conclusion, we have presented $\mu$SR measurements which demonstrate that
URu$_{2-x}$T$_x$Si$_2$ (T = Os, Fe) display antiferromagnetic order. This order
persists down to low doping levels, with our Fe = 0.02 sample showing a lowered
magnetic volume fraction that may indicate coexistence of HO and AF in this
sample. Furthermore, the magnetic order persists down to Fe doping levels below
that expected by a chemical pressure argument, and for Os dopings representing
negative chemical pressure, which shows that the hidden order is very fragile
and can easily be destroyed by even isoelectronic doping. These measurements,
combined with the local moment - hybridization crossover temperature from
susceptibility, demonstrate that magnetic order in isoelectronic doping is
driven by changes in hybridization rather than purely structural changes.

\section{Acknowledgments}

We thank Dr. G.D. Morris, Dr. B.S. Hitti and Dr. D.J. Arseneau (TRIUMF) for
their assistance with the $\mu$SR measurements. Work at McMaster university was
supported by the Natural Sciences and Engineering Research Council of Canada and
the Canadian Foundation for Innovation. M.N.W acknowledges support from the
Alexander Graham Bell Canada Graduate Scholarship program. T.J.W. acknowledges
support from the Wigner Fellowship program at Oak Ridge National Laboratory.
A.M.H acknowledges support from the Vanier Canada Graduate Scholarship program.
G.M.L. acknowledges support from the Canadian Institute for Advanced Research.
The Columbia University group acknowledges support from NSF DMR-1436095 (DMREF)
and OISE-0968226 (PIRE), JAEA Reimei project, and Friends of Univ. of Tokyo Inc.

\end{document}